\documentclass{JHEP3}
\usepackage{graphicx,amsmath}
\usepackage{dcolumn}% Align table columns on decimal point
\usepackage{bm}% bold math

\newcommand{\beq} {\begin{equation}}
\newcommand{\eeq} {\end{equation}    }
\newcommand{\bea} {\begin{eqnarray} }
\newcommand{\eea} {\end{eqnarray}    }

\newcommand{\sq}{\sqrt{2}}
\newcommand{\idthree}{{\mathbb I \/}_3}
\newcommand{\lm}{\lambda}
\newcommand{\Lm}{\Lambda}
\newcommand{\no}{\nonumber}

\newcommand{\Gm}{\Gamma}
\newcommand{\sg}{\sigma}

\newcommand{\gev}{~{\rm GeV}}
\newcommand{\tev}{~{\rm TeV}}

\newcommand{\sbt}{s_{\beta}}
\newcommand{\cbt}{c_{\beta}}
\newcommand{\tbt}{t_{\beta}}
\newcommand{\sbcb}{s_{\beta} c_\beta}
\newcommand{\xlm}{x_{\lambda}}

\newcommand{\lsim}{\mathrel{\mathop{\kern 0pt \rlap
  {\raise.2ex\hbox{$<$}}}
  \lower.9ex\hbox{\kern-.190em $\sim$}}}
\newcommand{\gsim}{\mathrel{\mathop{\kern 0pt \rlap
  {\raise.2ex\hbox{$>$}}}
  \lower.9ex\hbox{\kern-.190em $\sim$}}}

\title{Light pseudoscalar $\eta$ and $H \to\eta\eta$ decay\\
in the simplest little Higgs mode}

\author{Kingman Cheung \\
Department of Physics, National Tsing Hua University,
Hsinchu, Taiwan, R.O.C. \\
The National Center for Theoretical Sciences, Hsinchu, Taiwan \\
Email: {\tt cheung@phys.nthu.edu.tw} }

\author{Jeonghyeon Song \\
Department of Physics, Konkuk University, Seoul 143-701, Korea \\
Email: {\tt jhsong@konkuk.ac.kr} }

\preprint{hep-ph/0611294}

\abstract{ The SU(3) simplest little Higgs
model in its original framework without the so-called $\mu$
term inevitably involves a massless pseudoscalar boson $\eta$,
which is problematic for $b\,$-physics and cosmological axion limit.
With the $\mu$ term introduced by hand,
the $\eta$ boson acquires mass  $m_\eta \sim \mu$,
which can be lighter than half the Higgs boson mass
in a large portion of the parameter space.
In addition, the introduced $\mu$ term generates sizable coupling of
$H$-$\eta$-$\eta$.
The Higgs boson can dominantly decay into a pair of $\eta$'s
especially when $m_H$ below the $WW$ threshold.
Another new decay channel of $H \to Z \eta$
can be
dominant or compatible with $H \to W^+ W^-$
for $m_H$ above the $Z\eta$ threshold.
We show that the LEP bound on the Higgs boson mass is loosened
to some extent due to this new $H \to \eta\eta$ decay channel
as well as the reduced coupling of $H$-$Z$-$Z$.
The Higgs boson mass bound falls to about $110\gev$ for $f=3-4\tev$.
Since the $\eta$ boson decays mainly into a $b\bar b$ pair,
$H\to\eta\eta\to 4b$ and $H \to Z\eta \to Z b \bar{b} $
open up other interesting search channels in
the pursuit of the Higgs boson in the future experiments. We discuss
on these issues.}
\begin{document}

\section{Introduction}
\label{sec:introduction}

The Higgs boson is the last ingredient of the standard model (SM)
to be probed at experiments.
%The primary goal of the LHC is the search for the Higgs boson.
Precision measurements of the electroweak parameters
with logarithmic dependence on the Higgs boson mass
give indirect but tantalizing limit on $m_H$ to be less
than 186  GeV at the 95\% confidence level (C.L.)\cite{EWPD:mh}.
Direct search by the four LEP collaborations, ALEPH, DELPHI, L3 and OPAL,
resulted in no significant data.
A lower bound on the Higgs boson mass is established to be 114.4 GeV
at the 95\% C.L.\,\cite{LEP},
which is applicable to the SM and its extensions that preserve the
nature of the SM Higgs boson, \textit{e.g.}, minimal supersymmetric SM (MSSM)
in most parameter space.

In some other extensions, however, if  the nature of the light Higgs boson
is drastically modified,
the limit from direct search at LEP becomes weaker.
Phenomenologically, evading the LEP data is possible when the
Higgs boson coupling $g_{ZZH}$ with the $Z$ boson is reduced and/or
the Higgs boson decays into non-SM light particles.
In the CP-conserving MSSM, for example,
the lower bound on $m_H$ can be
in the vicinity of 93 GeV at the 95\% C.L.\,\cite{MSSM:Higgs:search}.
If we further allow CP violation
the result becomes more dramatic that
no absolute limits can
be set for the Higgs boson mass\,\cite{CPX_mh}.
Since the Higgs mass bound has far-reaching implications on the
Higgs search at the LHC,
the examination of the LEP bound on $m_H$ in other new models
is of great significance.

Recently, little Higgs models have drawn a lot of interests
as they can solve the little hierarchy problem
between the electroweak scale and the 10 TeV cut-off scale
 $\Lambda$\,\cite{LH}.
A relatively light Higgs boson mass
compared to $\Lm \sim 10\tev$ can be explained if % by the fact that
the Higgs boson is a pseudo-Nambu-Goldstone boson (pNGB) of
an enlarged global symmetry.
Quadratically divergent Higgs boson mass at one-loop level,
through the gauge, Yukawa, and self-couplings of the Higgs boson,
is prohibited by the collective symmetry breaking
mechanism.
%Phenomenologically, the one-loop level
%quadratic divergences from the SM gauge boson and top quark loops
%are canceled by those from new heavy gauge boson and heavy top
%loops, respectively.
%New heavy particles have masses at the symmetry breaking scale $f\sim 1\tev$.
According to the
global symmetry breaking pattern, there are various models with the
little Higgs mechanism\,\cite{various}.
Detailed studies have been also made, such as
their implications on electroweak precisions data (EWPD)\,\cite{EWPD}
and phenomenologies
at high energy colliders\,\cite{phenomenology}.

Considering the possibility of evading the LEP data on the Higgs mass,
the simplest little Higgs model\,\cite{simplest}
is attractive as it accommodates a light pseudoscalar boson $\eta$,
which the Higgs boson can dominantly decay into.
The model is based on [SU(3) $\times$ U(1)$_X]^2$ global symmetry
with its diagonal subgroup SU(3) $\times$ U(1)$_X$ gauged.
The vacuum expectation value (VEV) of
two SU(3)-triplet scalar fields,
$\langle\Phi_{1,2}\rangle = (0,0,f_{1,2})^T$,
spontaneously breaks both the global symmetry
and the gauge symmetry.
Here $f_{1,2}$ are at the TeV scale.
Uneaten pNGB's consist of a SU(2)$_L$ doublet
 $h$ and a pseudoscalar $\eta$.
Loops of gauge bosons and fermions generate
the Coleman-Weinberg  (CW) potential $V_{CW}$
which contains the terms such as $h^\dagger h$ and $(h^\dagger h)^2$:
The Higgs boson mass and its self-coupling are
radiatively generated.
However the CW potential with non-trivial operators of
$|\Phi_1^\dagger \Phi_2|^n$ does not have the dependence of $\eta$
which is only a phase of sigma fields $\Phi_{1,2}$\,\cite{CW,smoking}.
This $\eta$ becomes massless,
which is problematic for $\eta$ production in rare $K$ and $B$
decays, $\bar{B}$-$B$ mixing, and $\Upsilon \to \eta\gamma$,
as well as for the cosmological axion limit.

One of the simplest remedies was suggested
by introducing a $-\mu^2(\Phi_1 ^\dagger \Phi_2+ h.c.)$
term into the scalar potential by hand.
Even though this breaks
the global SU(3) symmetry and thus damages the little Higgs mechanism,
its contribution to the Higgs boson mass is numerically insignificant.
This $\mu$ determines the scale of $\eta$ mass.
By requiring negative Higgs mass-squared parameter for electroweak
symmetry breaking (EWSB),
we show that the $\mu$ (and thus $m_\eta$)
is of the order of 10 GeV.
Thus, we have light pseudoscalar particles.
In addition,
the $\mu$ term also generates the
$\lm' h^\dagger h \,\eta^2$ term in the CW potential.
As the $h$ field develops the VEV $v$,
$H$-$\eta$-$\eta$ coupling emerges with the strength proportional to $v\mu^2/f^2$,
with $f = \sqrt{f_1^2+ f_2^2}$ at the TeV scale.
The Higgs boson can then decay into two $\eta$ bosons.
Furthermore, this light $\eta$ opens a new decay channel
of $H \to Z\eta$.
Indeed, these two new decay channels can be dominant,
as shall be shown later.

Another issue which we make a thorough investigation into is the condition
for successful electroweak symmetry breaking (EWSB).
The model with the $\mu$ term is determined by four parameters: $f $,
$\tan\beta (= f_2/f_1)$, $\xlm$, and $\mu$.
Here $\xlm$ is the
ratio of two Yukawa couplings in the third generation quark sector.
The radiatively generated Higgs VEV $v$ is also determined by
these four parameters:
The SM EWSB condition $v =246\gev$ fixes one parameter, {\it e.g.,} $\tan\beta$.
For $\xlm \in [1,15]$, $\mu \sim \mathcal{O}(10)\gev$,
and $f=2-4\tev$,
the $v=246\gev$ condition limits $\tan\beta$ around 10.
This large $\tan\beta$
reduces the effective $g_{ZZH}$ coupling in this model.
With smaller $g_{ZZH}$ and $B(H\to b\bar b)$ than in the SM,
the LEP Higgs boson mass bound based on the limit $(g_{ZZH}/g^{\rm SM}_{ZZH})^2
\, B(H \to b\bar b)$ can be reduced \cite{LEP}.
Yet there was a general search by the DELPHI collaboration\,\cite{DELPHI}
in the channel $ e^+ e^- \to Z H \to Z (AA) \to Z + 4b$.  The $\eta$ boson
in the present model is similar to the $A$ boson.
We shall apply the limit obtained in the DELPHI analysis to the present
model, which shall be shown entirely unconstrained.

The organization of the paper is as follows. In the next section, we
highlight the essence of the original SU(3) simplest little Higgs model, in
particular the Higgs sector. We will show that the original model can
accommodate proper EWSB as well as the  Higgs mass $\sim 100\gev$.
After explicit demonstration of no $\eta$ dependence on the scalar potential,
we will discuss the problem of the massless pseudoscalar $\eta$.
In Sec.\ref{sec:mu}, we introduce the $\mu$ term and discuss the EWSB implication
as well as the mass spectra of the Higgs boson and $\eta$.
In Sec. \ref{sec:BR}, we calculate the branching ratio $H \to \eta \eta$ and
discuss its impact on the Higgs boson mass bound.
We discuss further possibilities to investigate this scenario and then
conclude in Sec. \ref{sec:Conclusions}.

\section{SU(3) simplest group model without the $\mu$ term}
\label{sec:simplest:nomu}
The SU(3) simplest little Higgs model is based on
$[\,$SU(3) $\times$ U(1)$_X]^2$ global symmetry
with its diagonal subgroup SU(3) $\times$ U(1)$_X$ gauged.
The pNGB multiplet is parameterized by two complex SU(3) triplet
scalar fields
$\Phi_{1,2}$:
\beq
    \Phi_1 = e^{i t_\beta \Theta} \Phi_{1}^{(0)}
        , \quad
    \Phi_2 = e^{-i \Theta/t_\beta } \Phi_{2}^{(0)}
        ,
\eeq
where $\tbt\equiv\tan\beta$ and
\begin{equation}
    \Theta = \frac{1}{f} \left[
        \left( \begin{array}{cc}
        \begin{array}{cc} 0 & 0 \\ 0 & 0 \end{array}
            & h \\
        h^{\dagger} & 0 \end{array} \right)
        + \frac{\eta}{\sqrt{2}}
        \left( \begin{array}{ccr}
        1 & 0 & 0 \\
        0 & 1 & 0 \\
        0 & 0 & 1 \end{array} \right) \right]
        \equiv \frac{1}{f} \,\mathbb{H}  +\frac{\eta}{\sq f} \idthree .
\end{equation}
The kinetic term for $\Phi_{1,2}$ is
\begin{equation}
\label{eq:LgPhi}
    \mathcal{L}_{\Phi} = \sum_{i=1,2} \left| \left( \partial_{\mu}
    + i g A_{\mu}^a T^a - \frac{i g_x}{3} B_{\mu}^x \right)
    \Phi_i \right|^2,
\end{equation}
where $T^a$
are the SU(3) generators while $A^a_\mu$ and $B_\mu$ are the SU(3) and U(1)
gauge fields, respectively.
Two gauge couplings of $g$ and $g_x$ are fixed by the SM gauge couplings
such that
SU(3) gauge coupling $g$ is just the SM SU(2)$_L$ gauge coupling
and $ g_x = {g'}/{\sqrt{1 - t_W^2/3}}$.

Each of the SM fermionic doublets is promoted
to a SU(3) triplet.
Focusing on the third generation quarks,
we introduce a $\mathbf{3}$ representation of SU(3),
$\chi_{L} = (t_{L},b_{L},iU_{L})^T$,
as well as two weak-singlet quarks, $U_{R1}$ and $U_{R2}$.
The Yukawa
interaction is
\beq
\label{eq:Yuk3}
{\cal L} =
i \lambda_1 U^{\dagger }_{R1} \Phi_1^\dagger \chi_L \,
+\,i \lambda_2 U^{\dagger} _{R2} \Phi_2^\dagger \chi^{}_L \,+\, {\rm h.c.},
\eeq
where the complex number $i$'s guarantee positive masses for fermions.
According to the SU(3) representation of the first two
generation quarks and all generation leptons,
there are two versions
for fermion embedding.
This variation in model building is possible
since light quarks and leptons
make very little contributions to the radiative Higgs mass.
The first fermion embedding is called  ``universal" embedding\,\cite{smoking},
where all three generations have identical quantum numbers.
The other is the ``anomaly-free"
embedding where anomaly-cancellation is required for easier UV
completion\,\cite{kong}:
The third generation quarks and all leptons are put into {$\mathbf{3}$} representations of SU(3),
while the first two generation quarks into {$\mathbf{\bar 3}$}.
Yukawa couplings for light quarks and leptons in both embedding cases
are referred to Ref.~\cite{smoking}.

When $\Phi_{1}$ and $\Phi_2$ develop the aligned VEV of
\beq
\langle \Phi_{1} \rangle =
\Phi_1^{(0)}=
(0,0,f\cos\beta)^T,
\quad
\langle \Phi_{2} \rangle = \Phi_2^{(0)}=
(0,0,f\sin\beta)^T,
\eeq
two kinds of symmetry breaking occur.
First,
 the global symmetry is spontaneously broken into its subgroup of
$[\,$SU(2) $\times$ U(1)$]^2$,
giving rise to ten Nambu-Goldstone bosons.
Second, the gauge symmetry SU(3) $\times$ U(1)$_X$ is broken into the SM
SU(2)$_L \times$ U(1)$_Y$, as five Nambu-Goldstone bosons are eaten.
Five new gauge bosons and one heavy top-like quark $T$ appear
with heavy mass of order $f \sim \tev$.
The heavy gauge bosons include
a $Z'$ gauge boson (a linear combination of $A^8_\mu$ and $B^x_\mu$)
and a complex SU(2) doublet $(Y^0,X^-)$
with masses of
\beq
M_{Z'}=\sqrt{\dfrac{2}{3-t_W^2}}\,g\, f, \quad M_{X^\pm}=M_Y=\dfrac{g f}{\sqrt{2}}
\,.
\eeq
The new heavy $T$ quark mass is
\beq
 M_T %= \sqrt{\lm_1^2 \cbt^2+\lm_2^2 \sbt^2} \,f
=\sqrt{2} \frac{\tbt^2+\xlm^2}{(1+\tbt^2)\xlm} \,\frac{m_t}{v} \,f
\,,
\eeq
where $\xlm = \lm_1/\lm_2$.

Brief comments on the EWPD constraint on $f$ are in order here.
According to Ref.~\cite{simplest},
the anomaly-free model is less constrained.
The strongest bound comes
 from atomic parity violation with $f>1.7\tev$ at the 95\%
C.L.
A more recent analysis in Ref.~\cite{Marandella:2005wd}
gives a stronger bound of $f>4.5\tev$ at 99\% C.L.
Main contribution comes from an oblique parameter $\hat{S}$
due to the $Z'$ gauge boson.
They applied the approximation for $Z'$ that is eliminated by solving its equation of motion.
Considering both analyses, we take $f=2-4\tev$ as reasonable choices.

The gauge and Yukawa interactions of the Higgs boson explicitly break the SU(3) global
symmetry,
generating the Higgs mass at loop level.
In the CW potential
up to dimension four operators,
only the $|\Phi_1^\dagger \Phi_2|^2$ term
leads to non-trivial result for the pNGB's.
A remarkable observation is that this
$|\Phi_1^\dagger \Phi_2|^2$ term does not have any dependence on
$\eta$\cite{Dias}.
This can be easily seen by the expansion of, \textit{e.g.}, $\Phi_{1}$ as
\beq
\label{eq:Phi12:matrix}
\Phi_1 = \exp\left( i\frac{\tbt\eta}{\sq f}\right)
\exp \left( i  \frac{\tbt }{f}\mathbb{H} \right)\Phi_{1}^{(0)},
\eeq
which we have used the Baker-Hausdorff formula
with $[\mathbb{H}, \idthree]=0$.
This compact form is very useful %.
%First the kinetic term is
%simplified as
%\beq
%\sum_{i=1,2}\left|\rd_\mu \Phi_i \right|^2 =
%\frac{1}{2}\rd_\mu \eta \, \rd^\mu\eta  + |\rd_\mu h|^2.
%\eeq
when calculating the $\Phi_1^\dagger \Phi_2$:
\beq
\label{eq:Phi:dagger:Phi2}
\Phi_1^\dagger \Phi_2 =
f^2 \sbt\cbt e^{-i \left(\tbt+\frac{1}{\tbt} \right) \frac{\eta}{\sq f} }
\cos \left(
 \frac{h_0}{f \cbt\sbt}
\right).
\eeq
The $|\Phi_1^\dagger \Phi_2|^2$ term
or the CW potential has no dependence on $\eta$.
Thus, the pseudoscalar $\eta$ remains massless in the original model.

On the contrary, the Higgs boson mass
is radiatively generated with one-loop logarithmic divergence
and two-loop quadratic divergence.
The troublesome one-loop quadratic divergence
is eliminated by the little Higgs mechanism.
The CW potential is
\begin{equation}
\label{eq:VCW:nomu}
    V_{\rm CW} = -m_0^2 \, h^\dagger h +\lm_0 (h^\dagger h)^2
,
\end{equation}
where
\begin{eqnarray}
\label{eq:msq}
    m_0^2 &=& \frac{3}{8 \pi^2} \left[
    \lambda_t^2 M_T^2 \ln \frac{\Lambda^2}{M_T^2}
    - \frac{g^2}{4} M_X^2 \ln \frac{\Lambda^2}{M_X^2}
    - \frac{g^2}{8} (1 + t_W^2) M^2_{Z^{\prime}}
    \ln \frac{\Lambda^2}{M^2_{Z^{\prime}}} \right],
 \\ \label{eq:lambda}
    \lambda_0 &=& \frac{1}{3 s_{\beta}^2 c_{\beta}^2} \frac{m_0^2}{f^2}
%\\ \no &&
    + \frac{3}{16 \pi^2} \left[
    \lambda_t^4 \ln\frac{M_T^2}{m_t^2}
    - \frac{g^4}{8} \ln \frac{M_X^2}{m_W^2}
    - \frac{g^4}{16} (1 + t_W^2)^2
    \ln \frac{M^2_{Z^{\prime}}}{m^2_Z} \right]\,.
\end{eqnarray}
Here $\lm_t = \sqrt{2}m_t/v$ and $\Lambda \simeq 4 \pi f$.
The negative mass-squared term for the Higgs doublet in Eq.\,(\ref{eq:VCW:nomu})
generates the VEV for the Higgs boson as
$\langle h \rangle  =v_0 /\sqrt{2}$,
which then triggers the EWSB and generates the Higgs boson mass $m_{H0}$,
given by
\bea
\label{eq:v:mH:meta}
v_{0}^2 = \frac{m_0^2}{\lm_0}, \quad m_{H0}^2 = 2 m_0^2
\,.
\eea

This CW potential alone has been considered
insufficient to explain the EWSB,
due to excessively large soft mass-squared $m_0^2$.
If $f=2\tev$ and $\xlm=\tbt=2$, for example,
$m_0\simeq 710\gev$ and thus $m_H \simeq 1\tev$.
In addition, the quartic coupling
$\lm_0$ is also small since it is generated by logarithmically divergent diagrams,
not by quadratically divergent ones.
In the ordinary parameter space of $\tbt$ and $\xlm$ of the order of one,
the $v_{CW} \simeq 246\gev$ condition cannot be satisfied.
However, this flaw
in the original model without the $\mu$ term
is not as serious as
usually considered in the literatures.
If we extend the parameter space allowing $\xlm$ and $\tbt$ up to $\simeq 10$,
the $v_{0} \simeq 246\gev$ condition can be met easily.
Reducing $m_0^2$ in Eq.~(\ref{eq:msq}) is possible
if the heavy $T$ mass decreases.
As discussed in Ref.~\cite{SingleT}, the heavy $T$ mass is minimized when $\tbt=\xlm$
and $\tbt$ increases.
Larger $\tbt$ can help to satisfy $v_0 \simeq 246\gev$.
In addition,
large $\tbt$ suppresses the new contributions to the EWPD\,\cite{SingleT}.

When we require that the radiatively generated Higgs VEV be equal
to the SM Higgs VEV,
what is the SM Higgs VEV in this model is an important question.
A definite way is to require that the SM Higgs VEV $v$ should explain
the observed SM $W$ gauge boson mass.
In this model, the $W$ gauge boson mass is modified into
\beq
\label{eq:mW}
m_W
=\frac{g v}{2} \left[
1-\frac{v^2}{12 f^2} \frac{\tbt^4-\tbt^2+1}{\tbt^2} + \mathcal{O} \left( \frac{v^4}{f^4}\right)
\right].
\eeq
The Higgs boson VEV explaining $m_W$, which we denote by $v_W$,
is
\beq
\label{eq:v}
v = v_0
\left[ 1+ \frac{v_0^2}{12 f^2}\frac{\tbt^4-\tbt^2+1}{\tbt^2}
+ \mathcal{O} \left( \frac{v^4}{f^4}\right)\right]
\equiv v_W
\,,
\eeq
where $v_0 = 2 m_W/g = 246.26 \gev$.
With the observed $m_W$,
the $v_W$ in this model depends on $\tbt$ and $f$.

\begin{figure}[th!]
\begin{center}
    \includegraphics[scale=0.8]{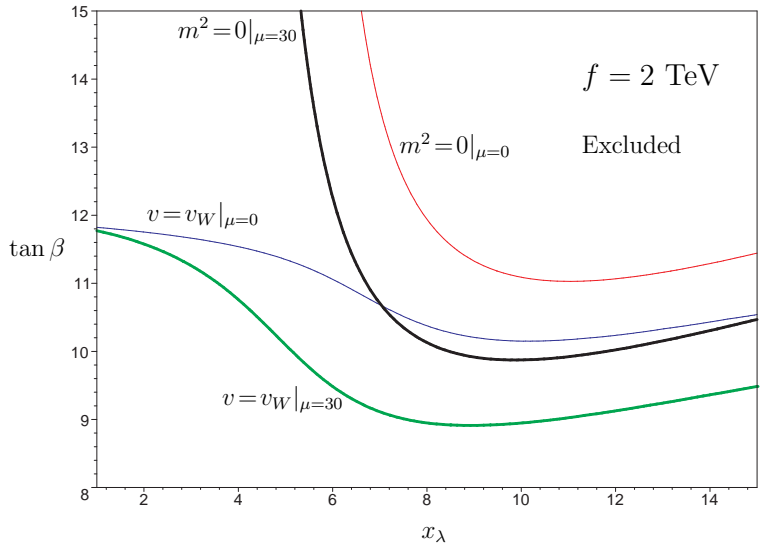}
\end{center}
    \caption {Allowed parameter space of $(x_\lambda,\; t_\beta)$ for $\mu=0,~30\gev$
     by valid
electroweak symmetry breaking. The red and blue (or thin) lines are
the contours of  $m^2=0$ and $v=v_W$ for $\mu=0$, respectively.
The black and green (or thick) lines satisfies
$m^2=0$ and $v=v_W$ for $\mu=30\gev$, respectively.
    }
    \label{fig:v-mu}
\end{figure}

In Fig. \ref{fig:v-mu},
we present the contours of
$m_0^2=0$ and $v_0=v_W$ (lines  for $\mu=0$).
In the upper right corner, $m^2_0$ becomes negative such that the EWSB is not
possible.  This is because too large $\tbt$ and thus too small
$M_T$ makes $m_0^2$ negative.
%%
%The $m_0^2=0$ contour appears as too large $\tbt$ reduces $M_T$ too much
%so that $m_0^2$ becomes negative.
%The upper right region is excluded since it cannot support EWSB.
%%
The $\lm_0 <0$ region is contained in the excluded region by $m_0^2=0$.
Thin lines are for $\mu=0$ case:
We do have considerably large parameter space, particularly around
$\tbt \simeq 10$, to explain appropriate EWSB.

Apparently the EWSB condition does not really need the extra $\mu$ term
if we can take large
$\tbt$ around 10.
The most serious problem is
the presence of \emph{massless} pseudoscalar $\eta$.
Any term in the CW potential,
proportional to $|\Phi_i^\dagger \Phi_i|^n$ or $|\Phi_1^\dagger \Phi_2|^n$,
cannot accommodate the $\eta$ dependence.
Even though lower bounds on CP-odd scalar masses from
the $b$-physics signal\,\cite{Hiller} and cosmology\,\cite{PDG} are not very stringent,
any pseudoscalar particle should be massive:
The $\eta$ mass can be as low as $\mathcal{O}(100)$ MeV
from the $b$-physics signal such as rare $K$, $B$ and radiative $\Upsilon$ decays
with the $\eta$ in the final state, $B_s\to\mu^+ \mu^-$ and
$B$-$\bar{B}$ mixing;
the cosmological bound is also weak but finite, as low as 10 MeV.
We should, therefore, extend the model to cure this massless
pseudoscalar problem.

\section{SU(3) model with the $\mu$ term}
\label{sec:mu}
The simplest solution to the massless $\eta$ problem
as well as generically large $m_0^2$ problem is to introduce
a new term of $-\mu^2 (\Phi^\dagger_1 \Phi_2 + h.c.)$
into the scalar potential
by hand\,\cite{simplest,Kaplan:Schmaltz,Kilian:pseudo-scalar}.
Unfortunately, this explicitly breaks the global SU(3) symmetry.
The little Higgs mechanism is lost as the Higgs loop
generates the one-loop quadratically divergent corrections to the Higgs mass.
Since this correction is numerically insignificant,
we adopt this extension.

Since the new term can be written as
\beq
-\mu^2 (\Phi^\dagger_1 \Phi_2 + h.c.)
=
- 2 \mu^2 f^2 \sbt\cbt \cos\left( \frac{\eta}{\sq \sbt\cbt f} \right)
 \cos \left(
 \frac{\sqrt{h^\dagger h}}{f \cbt\sbt}
\right),
\eeq
the scalar potential becomes
\beq
\label{eq:VCW}
V = - m^2 h^\dagger h + \lm (h^\dagger h)^2 - \frac{1}{2} m_\eta^2 \eta^2
+\lm' h^\dagger h \eta^2 + \cdots,
\eeq
where
\beq
\label{eq:msq:lambda}
m^2 = m_0^2 - \frac{\mu^2}{\sbcb}, \quad
\lm =\lm_0 - \frac{\mu^2}{12\sbt^3 \cbt^3},
\quad
\lm' = - \frac{\mu^2}{4 f^2 \sbt^3 \cbt^3}.
\eeq
The Higgs VEV $v$, the Higgs mass $m_H$, and $\eta$ mass $m_\eta$
are then
\beq
\label{eq:vsq:mH:meta}
v^2 = \frac{ m^2}{\lm} ,
\quad
m_H^2 = 2 m^2 ,
\quad
m_\eta^2 = \frac{\mu^2}{\sbcb}
\cos\left(
\frac{v}{\sq f \sbcb}
\right).
\eeq

The CW potential as well as the masses of new heavy particles
depend on the following four parameters:
\beq
\label{eq:parameters}
f,\quad \xlm, \quad \tbt,\quad \mu\,.
\eeq
As before, the $v=v_W$ condition removes one parameter.
In Fig.~\ref{fig:v-mu}, we present the contours of $v=v_W$ for $\mu=30\gev$
and $f=2\tev$.
Increasing $\mu$ reduces
the allowed value of $\tbt$ by $\sim 10\%$.

\begin{figure}[th!]
\begin{center}
    \includegraphics[scale=0.6]{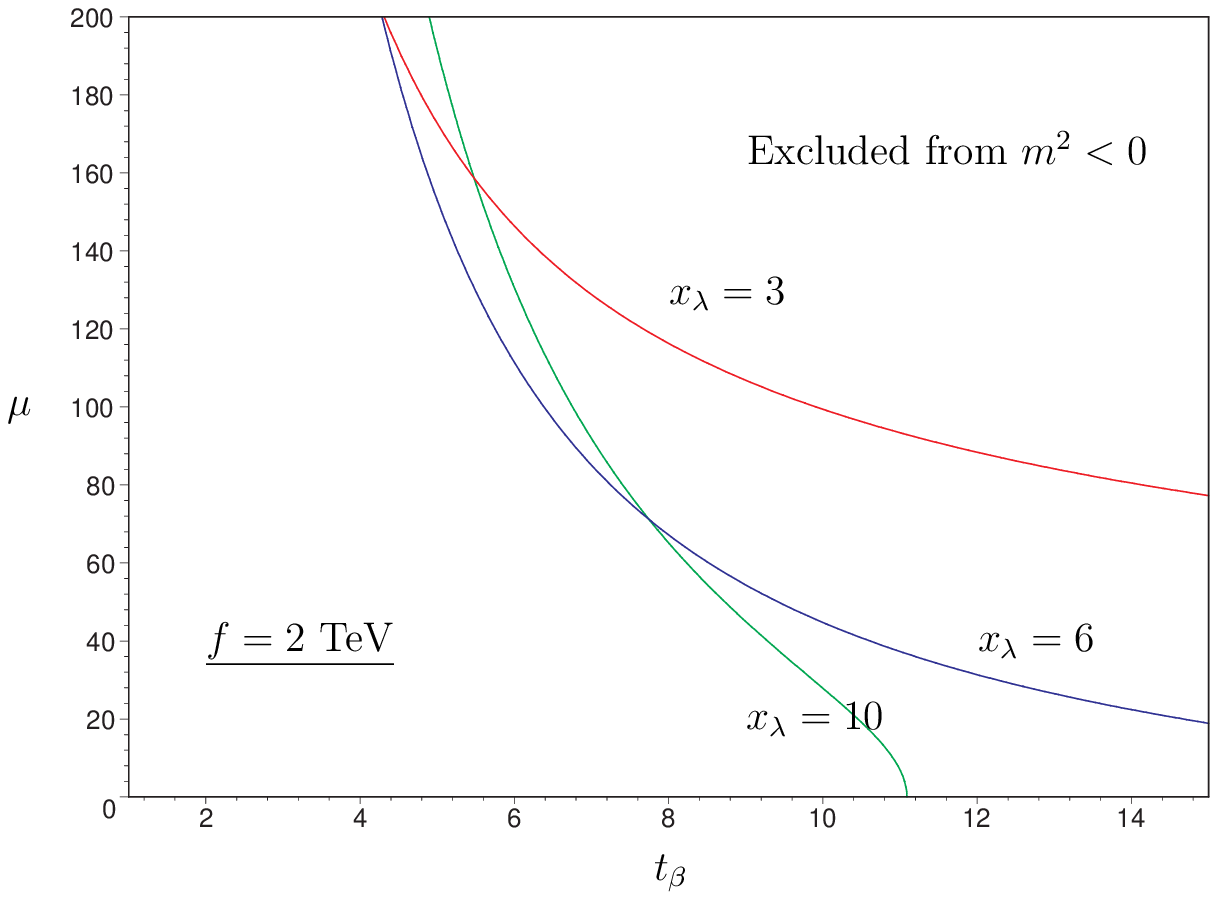}
    \includegraphics[scale=0.6]{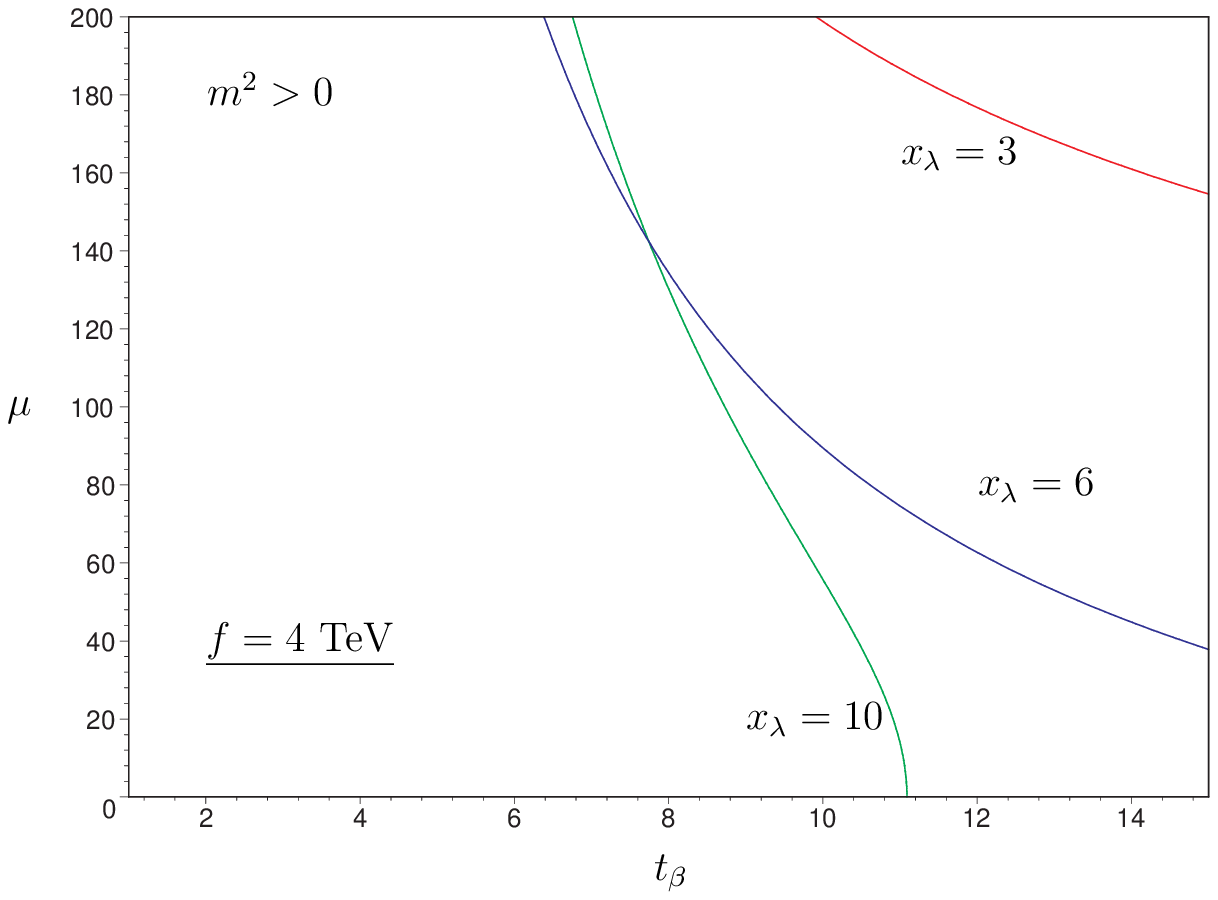}
\end{center}
    \caption {Allowed parameter space of $(\tbt,\; \mu)$ for $\xlm=3,~6,~10$
     by requiring positive Higgs mass-squared parameter $m^2$.
     We consider $f=2\tev$ and $f=4\tev$.
     Upper right corner is excluded since $m^2<0$.
    }
    \label{fig:mu:scale}
\end{figure}

Unfortunately there is no prior
information even about the scale of $\mu$.
Nevertheless upper bound on $\mu$ can be imposed
since $\mu$ contributes negatively to the Higgs mass-squared parameter $m^2$.
If $m^2$ becomes negative due to too large $\mu$,
the EWSB cannot occur.
In Fig.~\ref{fig:mu:scale},
we present the allowed parameter space of $(\tbt,\; \mu)$ for $\xlm=3,~6,~10$
and $f=2,4\tev$
by requiring $m^2>0$.
The upper right corner where $m^2 < 0$ is excluded due to the EWSB condition.
Since the $v=v_W$ condition prefers $\tbt\simeq 10$ as in Fig.~\ref{fig:v-mu},
the scale of $\mu$ is about $\mathcal{O}(10)$ GeV.

\begin{figure}[th!]
\begin{center}
    \includegraphics[scale=0.8]{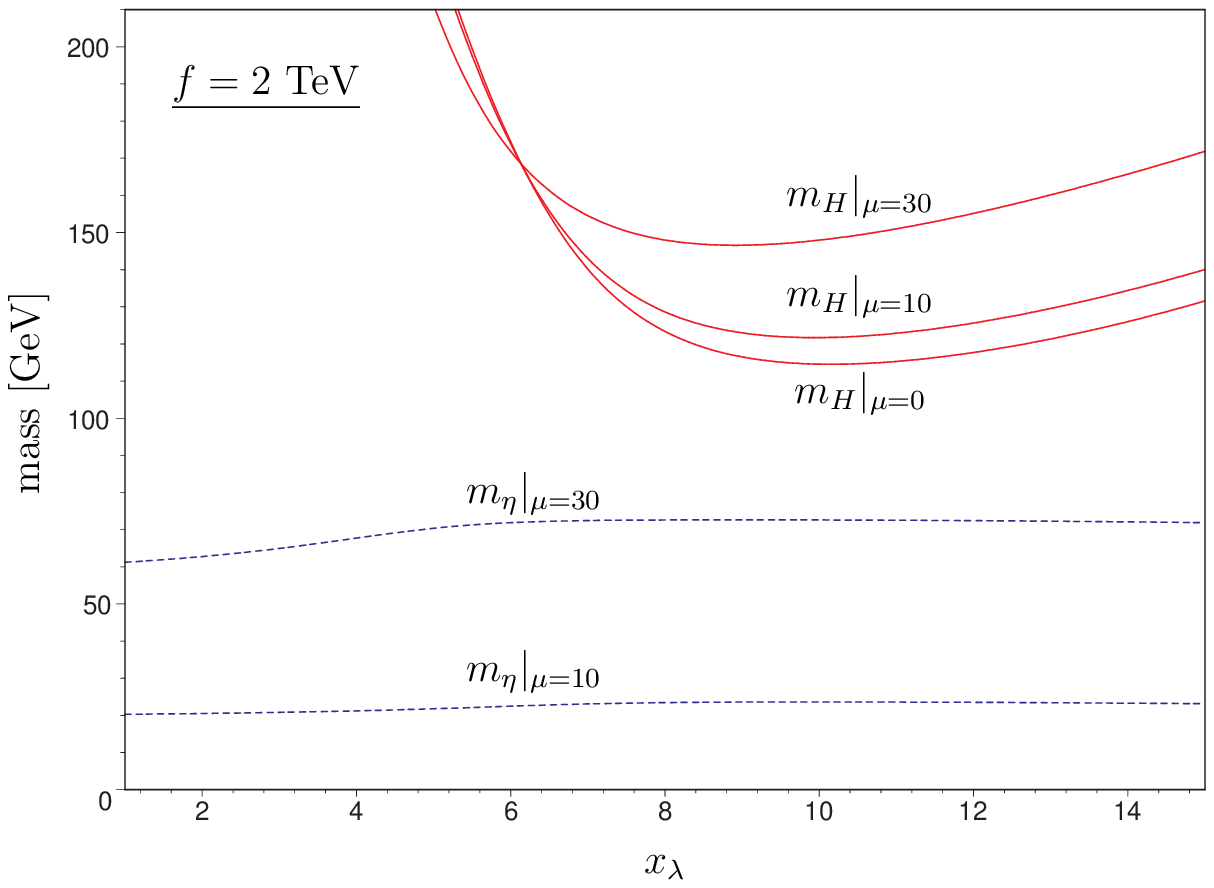}
\includegraphics[scale=0.8]{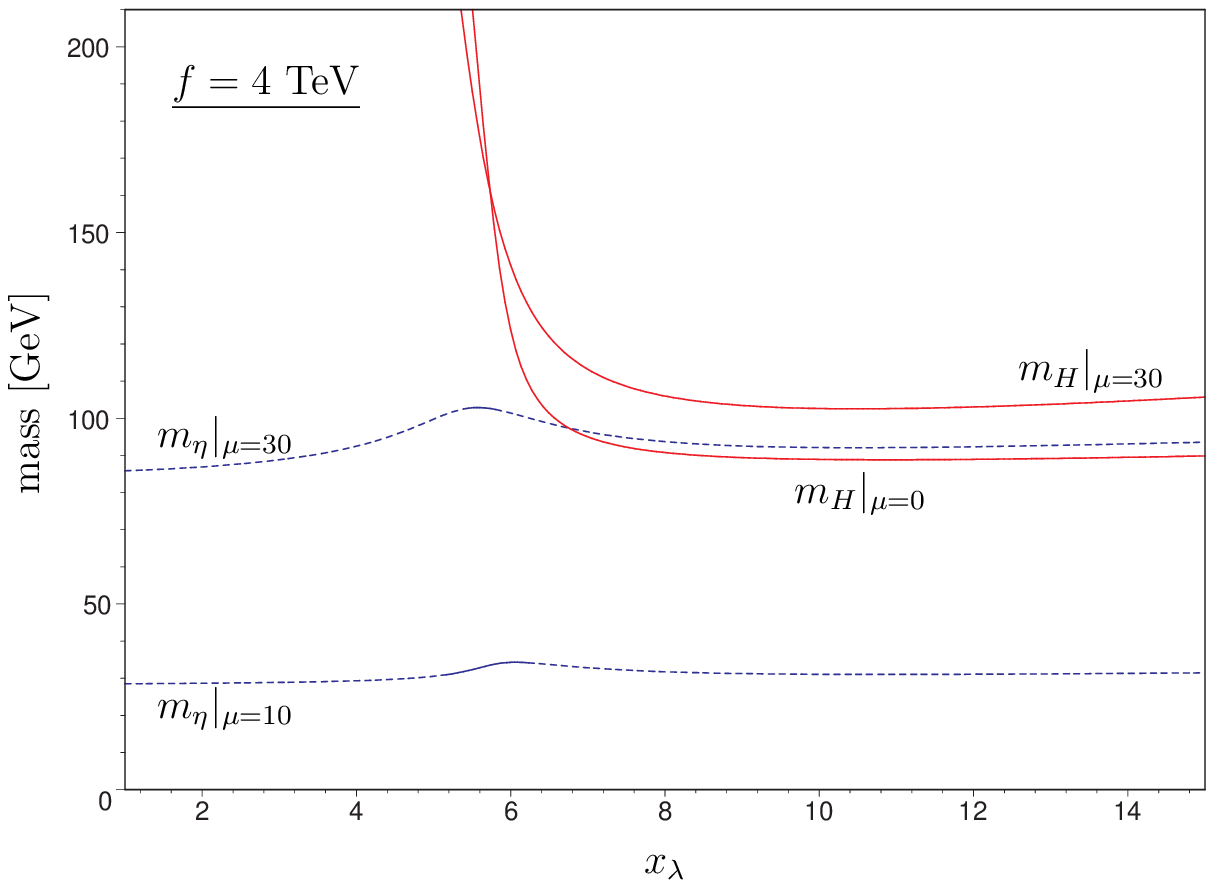}
\end{center}
    \caption {The masses of the Higgs boson (solid line) and $\eta$ (dashed line)
as a function of $\xlm$
    for $f=2\tev$  and $f=4\tev$.
    The value of $\tbt$ is determined by the $v_{CW}=v$ condition.
    }
    \label{fig:xlm:mheta}
\end{figure}

With the constraint of $v_{CW}=v$,
two parameters of $\xlm$ and $\mu$
determine the masses of the Higgs boson and $\eta$ at a given $f$.
In Fig.~\ref{fig:xlm:mheta},
we plot, as a function of $\xlm$, the $m_H$ (solid lines) and  $m_\eta$ (dashed line)
for $\mu=0,~10,~30\gev$ and $f=2,4\tev$.
Note that $m_\eta=0$ for $\mu=0$.
For non-zero $\mu$,
the $\eta$ mass is around $\sim \mathcal{O}(10)\gev$.
And the Higgs boson mass is generically around $\sim $100 GeV.

In addition, we find some other interesting features.
First, both $m_H$ and $m_\eta$ attain a minimum with a given $f$,
which occurs when $\mu=0$.
This minimum of the Higgs boson mass is close to the LEP bound of $114.4\gev$,
and decreases as $f$ increases.
For example, $m_H^{(\mathrm{min})}=114.5\gev$  for $f=2\tev$,
and $m_H^{(\mathrm{min})}=88.9 \,(88.8) \gev$ for $f=3\,(4)\tev$.
Investigation of the LEP bound on the Higgs boson mass is
of great significant in this model.
Second, $\mu$ increases both $m_H$ and $m_\eta$.
Since $m_\eta \propto \mu$ as in Eq.(\ref{eq:v:mH:meta}),
increasing $m_\eta$ with $\mu$ is easy to understand.
However $m_H$ has negative contribution from increasing $\mu$
as in Eq.~(\ref{eq:msq:lambda}):
Increasing $m_H$ with $\mu$ seems strange.
This behavior is due to the $\tbt$ value determined by the $v=v_W$ condition.
With high $\mu$, the $\tbt$ value for $v=v_W$
is reduced as in Fig.\ref{fig:v-mu}:
Smaller $\tbt$ raises the $M_T$, and thus also raises
its radiative contribution to
the Higgs boson mass.

\begin{figure}[th!]
\begin{center}
    \includegraphics[scale=0.8]{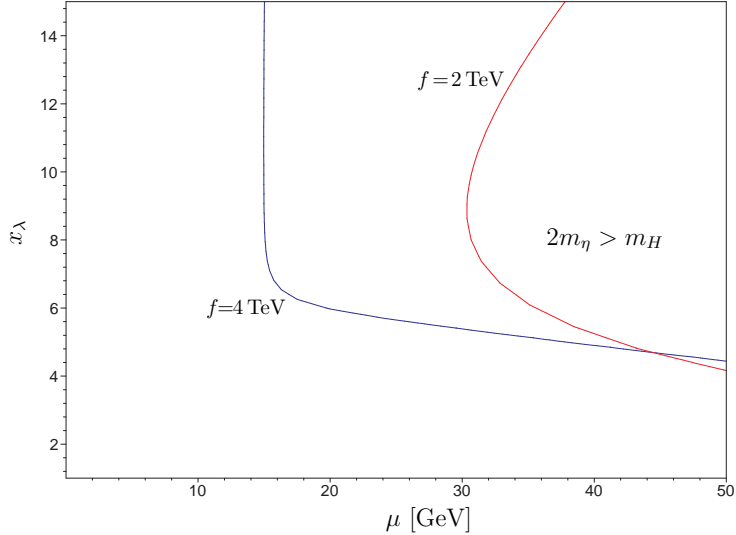}
\end{center}
    \caption {The contours of $m_H = 2 m_\eta$ in the parameter
space $(\mu,\xlm)$ for $f=2,4\tev$. To the left-hand (right-hand) side of
the contour, $2 m_\eta < \;( > )\, m_H$
    }
    \label{fig:mratio}
\end{figure}

Another important point is that $m_\eta$ can be quite light.
In principle,
$m_\eta$ can be as light as the current $b$ physics and/or cosmological bounds allow.
In this paper, however, we adopt the generic mass scale for $\eta$,
around $\mathcal{O}(10)\gev$.
This light pseudoscalar particle can have
a significant implication on the phenomenology of the Higgs boson.
The $\lm' h^\dagger h \eta^2$ term in the scalar potential of Eq.(\ref{eq:VCW})
leads to the coupling of $H$-$\eta$-$\eta$:
If $\eta$ boson is light enough, the Higgs boson can decay into a pair of $\eta$
and the Higgs discovery strategy should be reexamined.
In Fig.\ref{fig:mratio},
we present, with $f=2,4\tev$, the parameter space of $(\mu,\xlm)$
where $2 m_\eta < m_H$ (to the left-hand side of the contours).
If $\mu$ is too large, $H\to \eta\eta$ decay is kinematically prohibited
unless $\xlm$ is smaller than a certain value.

\section{$H \to \eta\eta$ Decay and LEP implications}
\label{sec:BR}
 \subsection{Branching ratios}
In this model, major decay modes of the Higgs boson
are SM-like ones with the partial decay rates as
\bea
\label{eq:Gamma}
\Gm(H \to f\bar{f}) &=& \frac{N_C g^2 m_f^2}{32 \pi m_W^2} (1-x_f)^{3/2} m_H,
\qquad \mbox{for $f=t,b,c,\tau$},
\\ \no
\Gm(H \to W^+W^-) &=& \frac{g^2}{64 \pi} \frac{m_H^3}{m_W^2}
  \sqrt{1-x_W} \left(1-x_W + \frac{3}{4}x_W^2 \right), \\ \no
\Gm(H \to ZZ) &=& \frac{g^2}{128 \pi} \frac{m_H^3}{m_Z^2}
  \sqrt{1-x_Z} \left(1-x_Z + \frac{3}{4}x_Z^2 \right),
\eea
where $x_i = 4 m_i^2/m_H^2$, $N_c = 3\,(1)$ for $f$ being a quark (lepton).
New decay channels are
\bea
\label{eq:Gamma:new}
\Gm(H \to \eta\eta) &=& \frac{{\lm'}^2}{8\pi}\frac{v^2}{m_H} \sqrt{1-x_\eta}
=\frac{m_\eta^4 }{8 \pi v^2 m_H}\sqrt{1-x_\eta}
,\\ \no
\Gamma( H \to Z \eta) &=& \frac{m_H^3}{32 \pi f^2}
  \left( t_\beta - \frac{1}{t_\beta} \right)^2 \,
  \lambda^{3/2} \left(1, \frac{m_Z^2}{m_H^2}, \frac{m_\eta^2}{m_H^2}
\right ),
\eea
where $\lambda (1,x,y) = (1-x-y)^2 - 4 xy$.
The last decay mode was mentioned in Ref. \cite{Kilian:pseudo-scalar},
which could be dominant and phenomenologically quite interesting.
%As shall be shown, when $m_H$ is sufficiently above the LEP bound,
%the $H \to Z \eta$ decay will become substantial if not dominant.

\begin{figure}[th!]
\begin{center}
    \includegraphics[scale=.58]{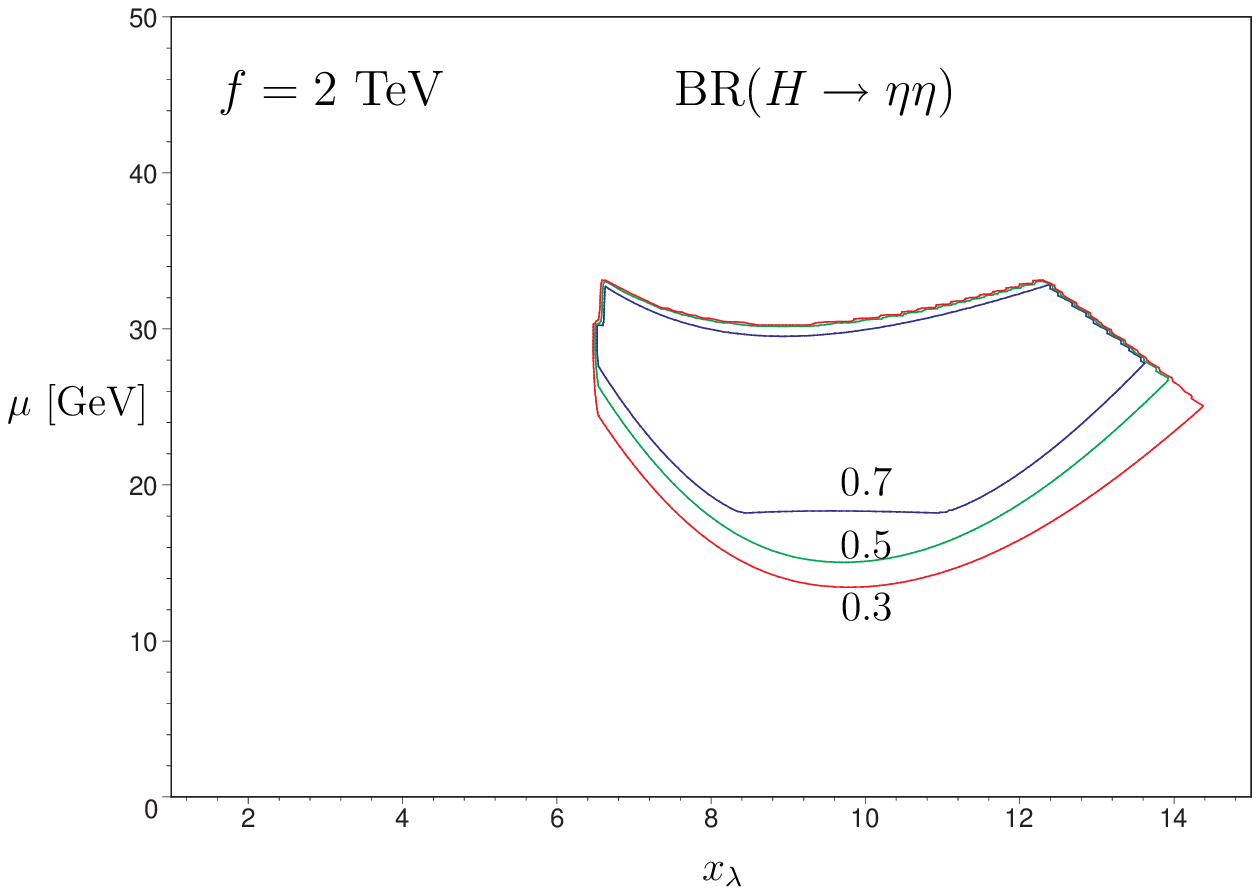}
\includegraphics[scale=.58]{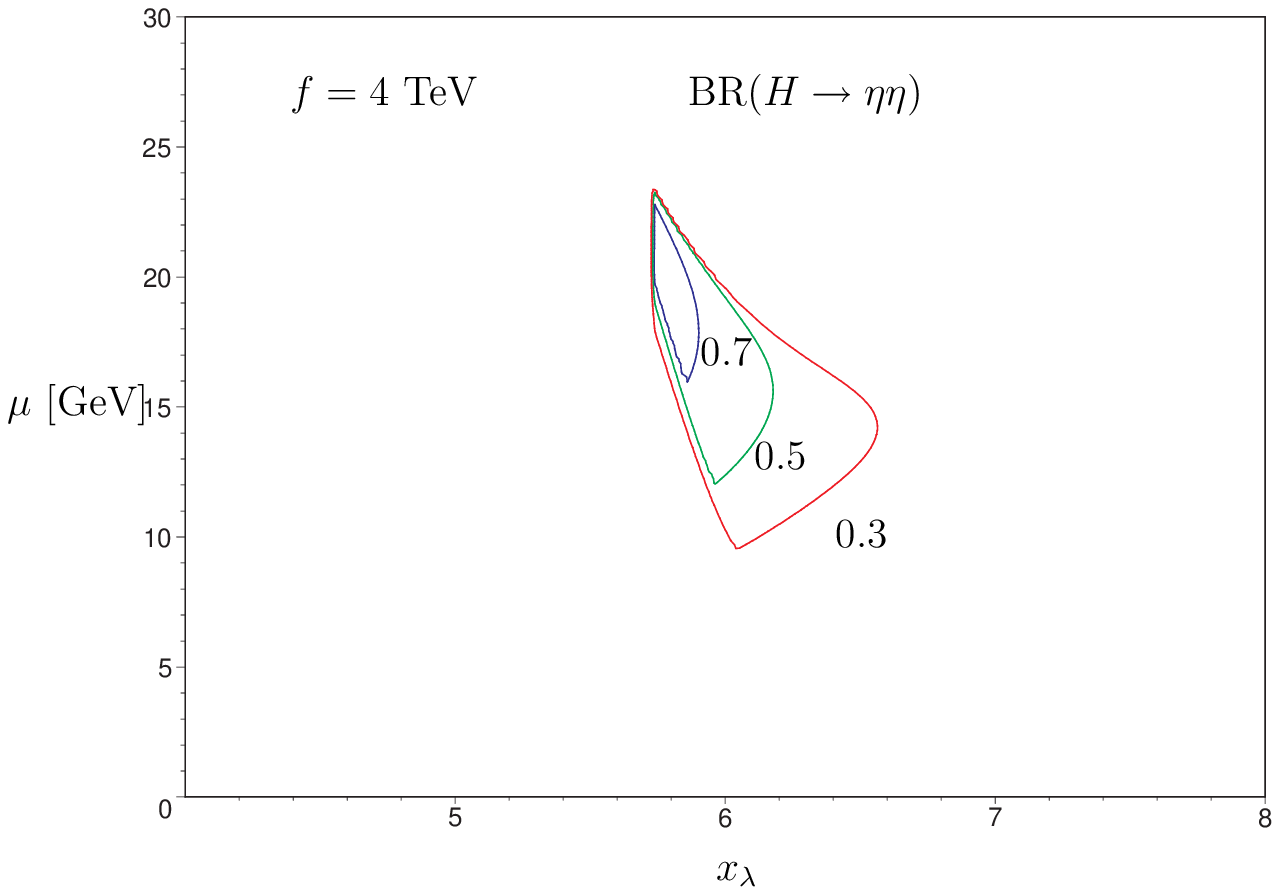}
\end{center}
    \caption {Contours of $B(H \to \eta \eta)=0.3,0.5,0.7$
    in the parameter space $(\xlm,\mu)$
    for $f=2,4\tev$.
    }
    \label{fig:BReta:contour}
\end{figure}

Search strategy of the Higgs boson depends sensitively
on its branching ratios (BR):
In the SM, the major decay mode for $m_H < 2 m_W$
is into $b\bar{b}$
while that for $m_H \gsim 2 m_W$ is into $W^+ W^-$.
In this model, there are two new decay modes for the Higgs boson,
$H \to \eta \eta$ and $H \to Z \eta$.
In Fig.~\ref{fig:BReta:contour}, we present the
contours of $B(H \to \eta \eta)=0.3,0.5,0.7$
in the parameter space $(\xlm,\mu)$
for $f=2,4\tev$.
%%
%
%The $f$-dependence on the contours are
%quite small.
%
Quite sizable portions of the parameter space can accommodate
dominant decay of $H \to \eta\eta$.
For $f=2\tev$,
$B(H \to \eta \eta)> 0.5$ requires $\xlm \in [6,14]$
and $\mu \in [16,30]$ GeV.
A smaller $\mu$ increases the 2-body phase-space factor
since $\mu$ is proportional
to the produced $\eta$ mass,
while it reduces the $H$-$\eta$-$\eta$ coupling.
The optimal $\mu$ for large $B(H \to \eta \eta)$
is around 20 GeV.
The size of parameter space for $f=4$ TeV is relatively smaller
with $\xlm \in [5.6,6.6]$ and $\mu \in [10,22]\gev$.  In this case, the
optimal $\mu$ is also around 20 GeV.

\begin{figure}[th!]
\begin{center}
    \includegraphics[scale=.58]{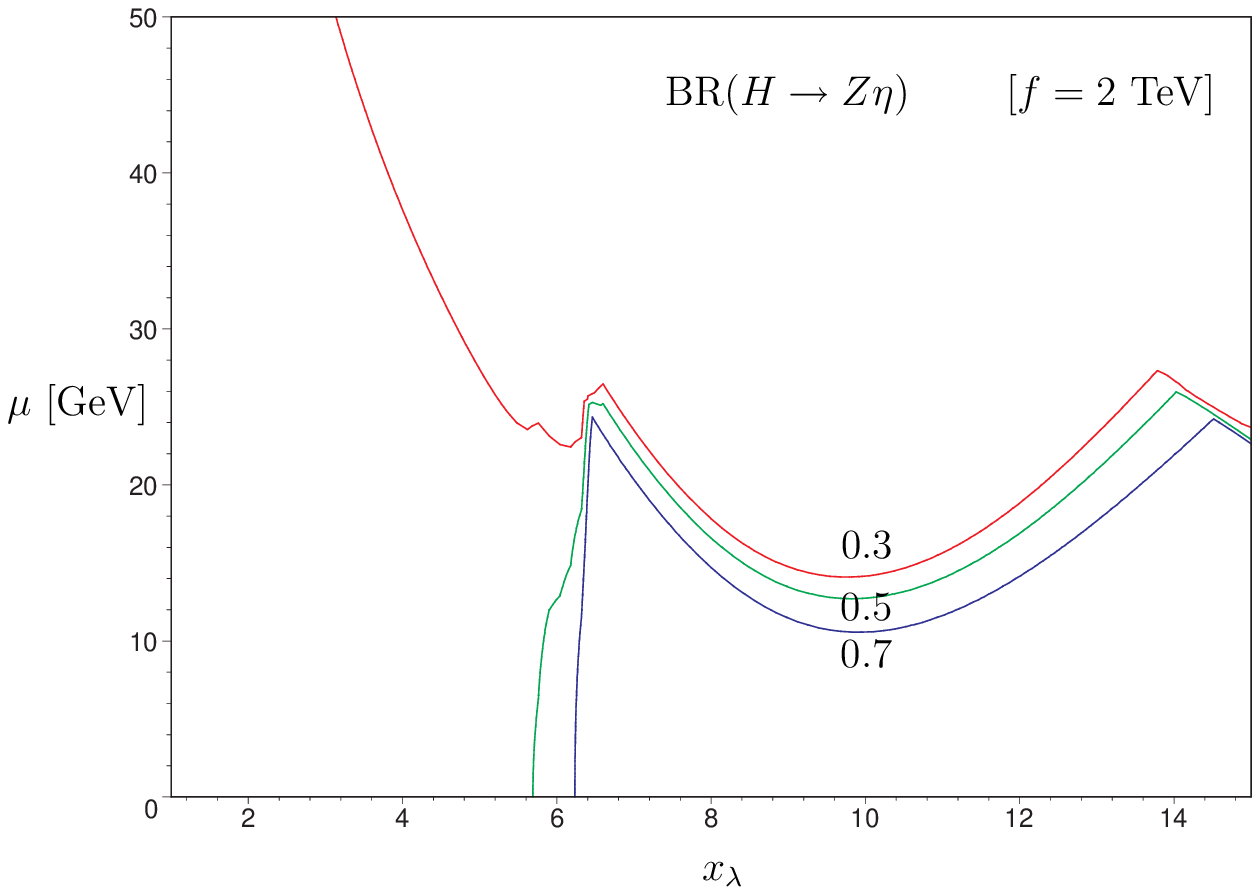}
\includegraphics[scale=.58]{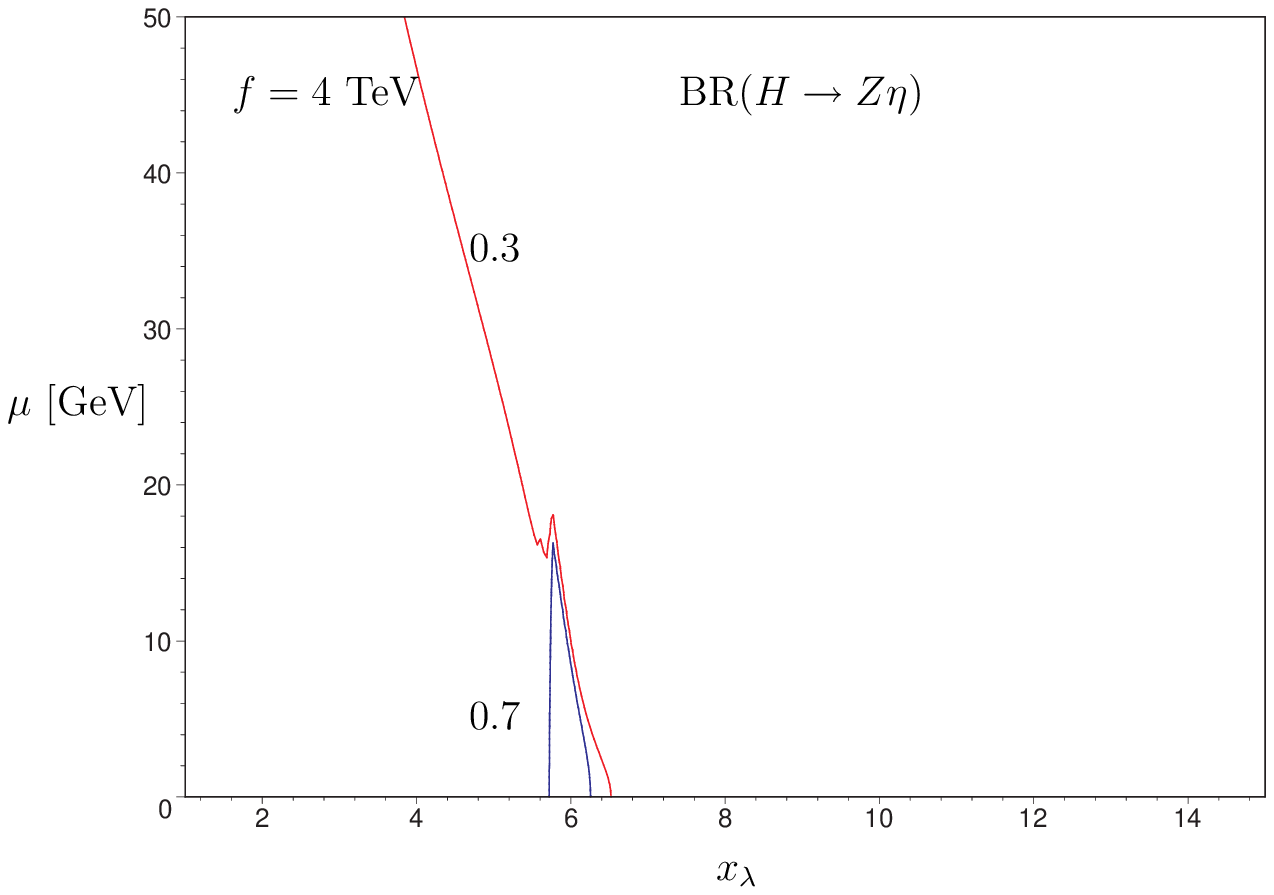}
\end{center}
    \caption {Contours of $B(H \to Z \eta)=0.3,0.5,0.7$
    in the parameter space $(\xlm,\mu)$
    for $f=2,4\tev$.
    }
    \label{fig:BRZeta:contour}
\end{figure}

Figure \ref{fig:BRZeta:contour} shows the same contours
for $B(H \to Z \eta)$,
which
depend quite sensitively on $f$.
For $f=2\tev$,
sizable parameter space of
$\xlm \gsim 6 $ and $\mu\lsim 10\gev$
can allow dominant decay of $H \to Z\eta$.
When $f=4\tev$,
only a small region around $\xlm \simeq 6$ and $\mu\lsim 15\gev$
can accommodate dominant $H \to Z\eta$.
This is mainly due to the $\eta$ mass.
As can be seen in Fig.\ref{fig:xlm:mheta},
$\eta$ for $f=4\tev$ is relatively heavier than that for $f=2\tev$.

\begin{figure}[th!]
\begin{center}
    \includegraphics[scale=.8]{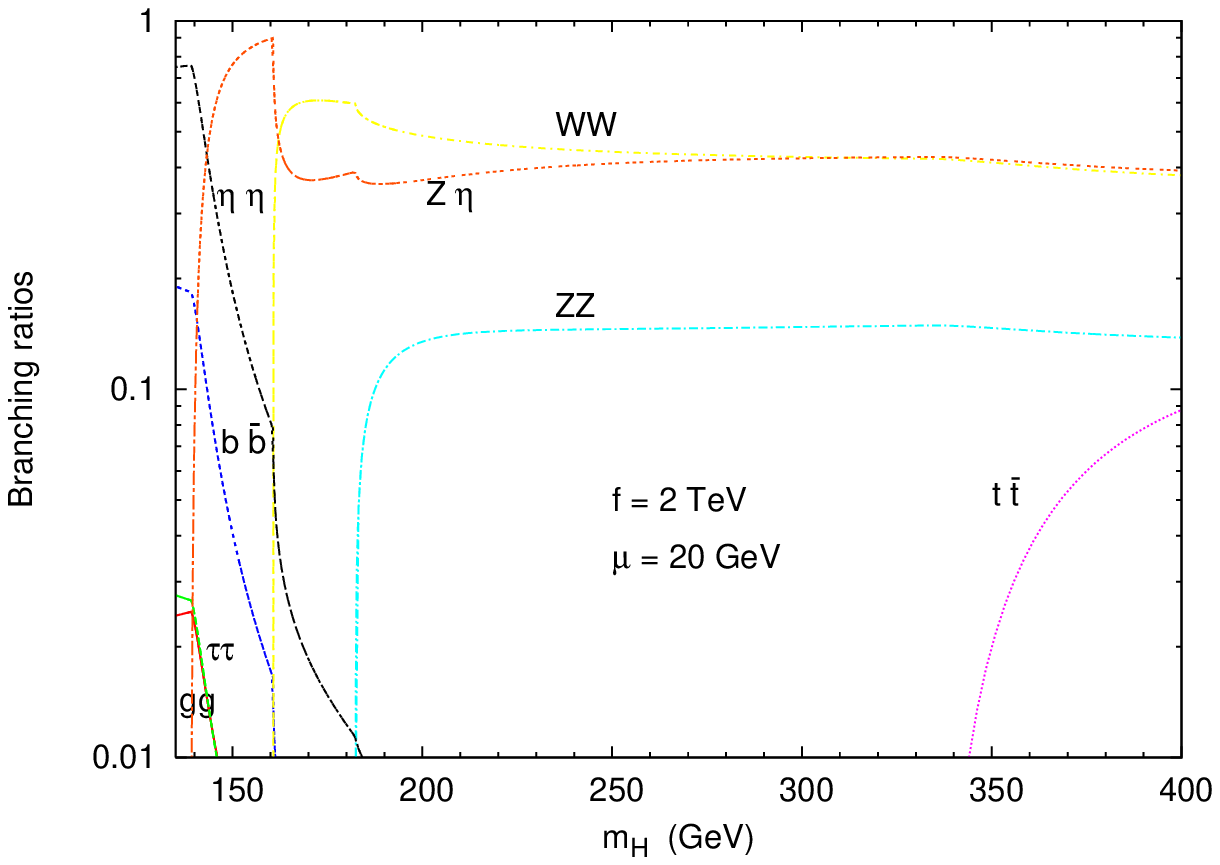}
\includegraphics[scale=.8]{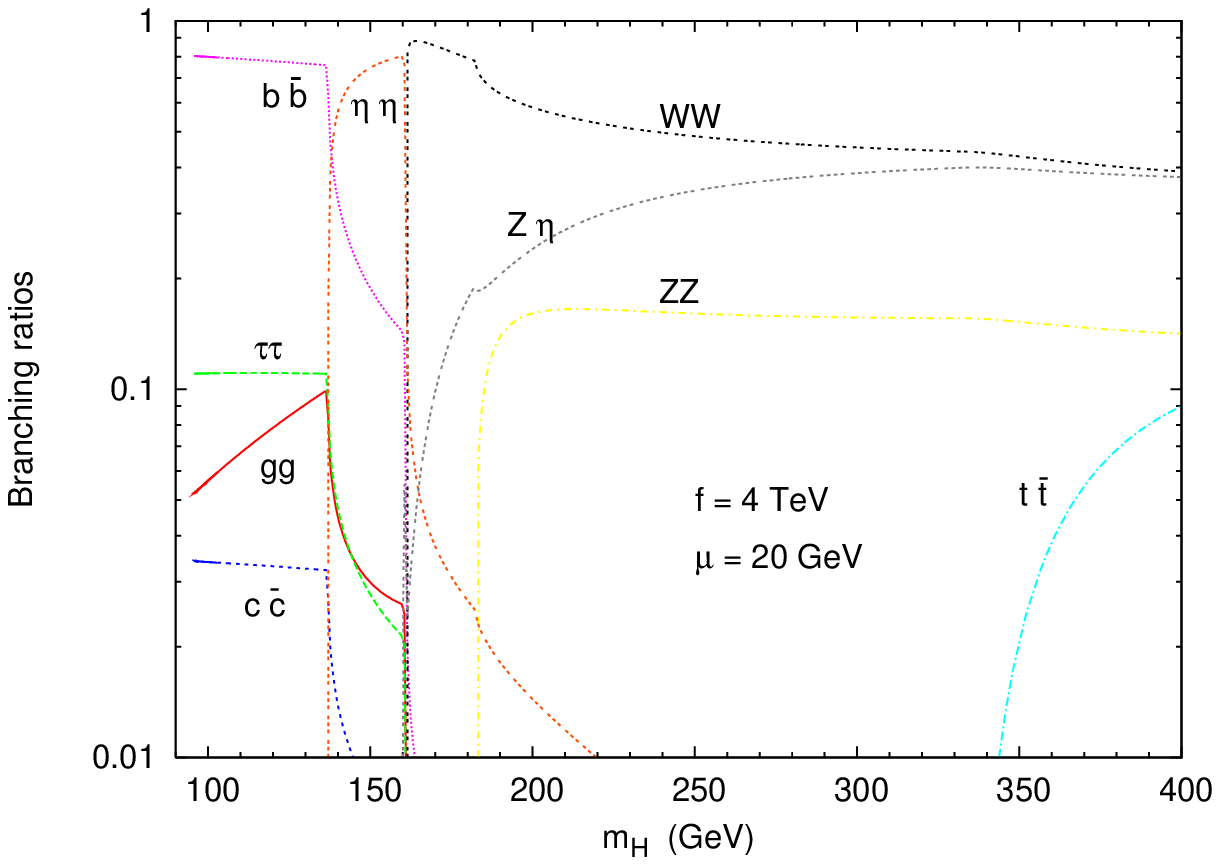}
\end{center}
    \caption {Branching ratios of the Higgs boson
in the simplest little Higgs model with the $\mu$ term
as a function of $m_H$ for $f=2\tev$ and $f=4\tev$.
We fix $\mu =20\gev$ but vary $\xlm$.
    }
    \label{fig:BR}
\end{figure}

In order to see the $m_H$ dependence on each branching ratio,
we present the branching ratios
as a function of $m_H$ for $f=2,4\tev$
in Fig.~\ref{fig:BR}.
We fix $\mu=20$ GeV for both $f=2,4$ TeV
while vary $\xlm$ to generate various $m_H$.
Different distribution of BRs for $f=2\tev$ from that for $f=4\tev$
is mainly due to the Higgs mass range.
In the $f=2\tev$ case,
$Z \eta$ mode is solely dominant for $m_H$ from the $Z\eta$ threshold
to $2m_W$.
Even for $m_H > 2 m_W$ $B(H \to Z\eta)$
is almost the same as $B(H \to W^+ W^-)$.
In the $f=4\tev$ case,
the $H \to \eta\eta$ is dominant for $140 \lsim  m_H \lsim 160$ GeV,
but the $H \to b\bar{b}$ becomes dominant if $m_H$ is below about 140 GeV.
For $m_H$ above $WW$ threshold,
$H \to W W$ is the leading decay mode, but not as dominant as in the SM
because of the presence of the $Z \eta$ mode.
The second important decay mode is into $Z \eta$,
which is very different from a
SM-like Higgs boson~\cite{Kilian:pseudo-scalar}.

Brief comments on the decay of $\eta$ is in order here.
If $m_\eta<2 \, m_W$, the decay pattern is very similar to that of the
SM Higgs boson
with the main decay mode into a SM fermion pair
via the coupling $c (m_f/f) i \bar f \gamma_5 f$, where
$c \sim O(t_\beta)$ and $m_f$ is the mass of the fermion.
Although this coupling is suppressed by $1/f$, the decay is still prompt
in collider experiments for $f \sim O({\rm TeV})$.
Therefore, the light $\eta$ boson mainly
decays into a $b \bar{b}$ pair\,\cite{Kilian:pseudo-scalar} if kinematically
allowed.
This characteristic feature of $\eta$ decay is useful to probe $\eta$
at high energy colliders.

\subsection{LEP bound on $m_H$}
\begin{figure}[th!]
\begin{center}
    \includegraphics[scale=1]{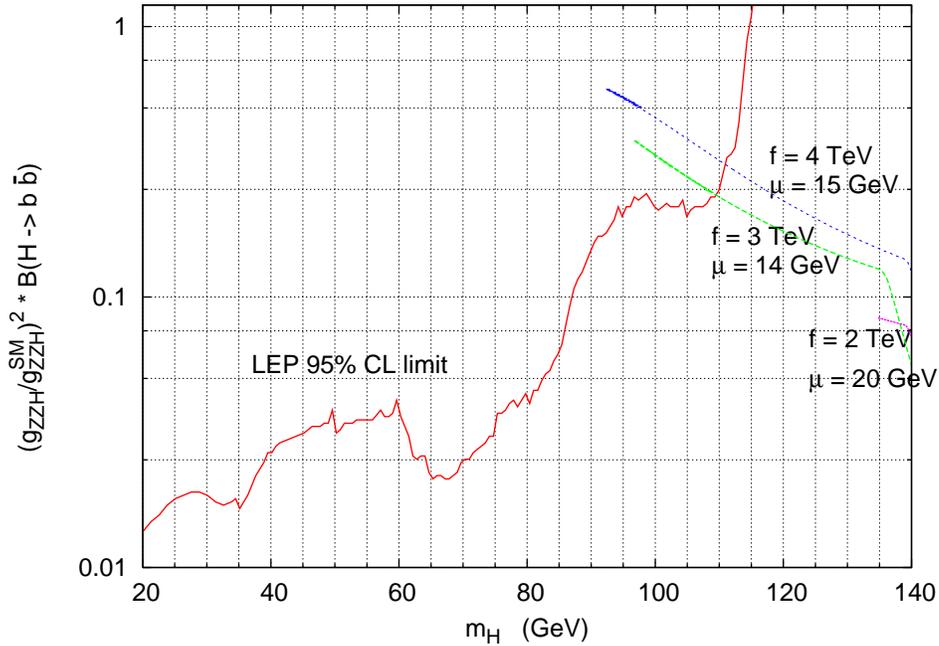}
\end{center}
  \caption {Upper bound on $[g_{ZZH}/g^{\rm SM}_{ZZH} ]^2 \times B( H \to b\bar b)$
established by the LEP collaborations,
and the corresponding values in the simplest little Higgs model
that we are considering.
    }
    \label{fig:LEP2}
\end{figure}

Due to the presence of dominant decay of $H \to \eta\eta$, one may
expect that the LEP bound on the Higgs mass can be loosened to some extent.
The four LEP collaborations~\cite{LEP} searched for the Higgs boson via
\beq
\label{eq:LEP:process}
e^+ e^- \to Z H \to (l^+ l^-,q\bar{q},\nu\bar{\nu}) + b\bar{b}.
\eeq
Here the main decay mode of the SM Higgs boson into $b\bar{b}$ dominates
the width of the Higgs boson, with a branching fraction about 90\% for
most of the mass range and down to about 74\% at $m_H= 115$ GeV.
There is also a search using a minor mode of $H \to \tau^+ \tau^-$.
Nevertheless, the combined
limit is almost the same as that using just the $b\bar b$ mode.
The mass bound on the SM Higgs boson is 114.4 GeV \cite{LEP}.
For model-independent limits the
LEP collaborations presented the upper bound on
$[g_{ZZH}/g^{\rm SM}_{ZZH} ]^2 \times B( H \to b\bar b)$
at the 95\% C.L., as shown by the rugged curve in
Fig. \ref{fig:LEP2}.
%Note that the SM limit can be recovered by setting
%$[g_{ZZH}/g^{\rm SM}_{ZZH} ]^2 \times B( H \to b\bar b) =1$.

In the simplest little Higgs scenario with the $\mu$ term,
one anticipates that the LEP bound on $m_H$ would be reduced, because of
(i) sizable decay rate of $H\to \eta\eta$ such that $B(H\to b\bar b)$
is substantially reduced as shown in Fig. \ref{fig:BR},  and
(ii) the reduced coupling $g_{ZZH}$ in the simplest little Higgs model, especially
when $t_\beta$ is large.
In this model,
the $g_{ZZH}$ deviates from the SM value by
\beq
\frac{g_{ZZH}}{g_{ZZH}^{\rm SM}}  =
\left[
1-\frac{v_0^2}{4 f^2}\left\{\tbt^2-1+\frac{1}{\tbt^2} +(1-t_W^2)^2 \right\}
\right]\,.
\eeq

In Fig.~\ref{fig:LEP2}, we present the prediction of
$[g_{ZZH}/g^{\rm SM}_{ZZH} ]^2 \times B( H \to b\bar b)$
for $f=2,3,4\tev$, and compare to the 95\% C.L. upper limit obtained
by the LEP collaborations.
We found the best value of $\mu = 14\;(15)$ GeV for $f=3\;(4)$ TeV
such that the prediction of
$[g_{ZZH}/g^{\rm SM}_{ZZH} ]^2 \times B( H \to b\bar b)$
for $f=2,3,4\tev$ is the smallest.
The $f=2\tev$ case is safe because the minimum value of $m_H$ predicted
is already above 114 GeV.
For $f=3,4\tev$, however,
the Higgs boson mass bound is restricted by the data as follows:
\bea
m_H &>& 109 \gev ~~ \quad   \mathrm{for}~~ f = 3 \tev, \\ \no
   m_H &>& 111 \gev~~\quad   \mathrm{for}~~ f = 4 \tev.
\eea

\subsection{DELPHI limit on $C^2_{Z(AA\to4b)}$}
The DELPHI collaboration~\cite{DELPHI} has searched for
the process $e^+ e^- \to ZH \to Z(AA) \to Z + 4b$
for $m_H>2 m_A$.
Here $A$ is a CP-odd scalar particle, for which $\eta$ is a good candidate.
The DELPHI collaboration parameterized the cross section by
\beq
\sg_{(AA)Z \to 4 b+jets} = \sg_{HZ}^{\rm SM} \times B(Z \to\mathrm{ hadrons}) \times
C^2_{Z(AA\to4b)},
\eeq
where
\beq
C^2_{Z(AA\to4b)} = \left( \frac{g_{ZZH}}{g_{ZZH}^{\rm SM}} \right)^2
\times
B(H\to A A) \times B(A\to b\bar{b})^2.
\eeq
As no convincing evidence for a signal was found,
the upper bound on $C^2_{Z(AA\to4b)}$ was presented~\cite{DELPHI}.

\begin{figure}[th!]
\begin{center}
    \includegraphics[scale=1]{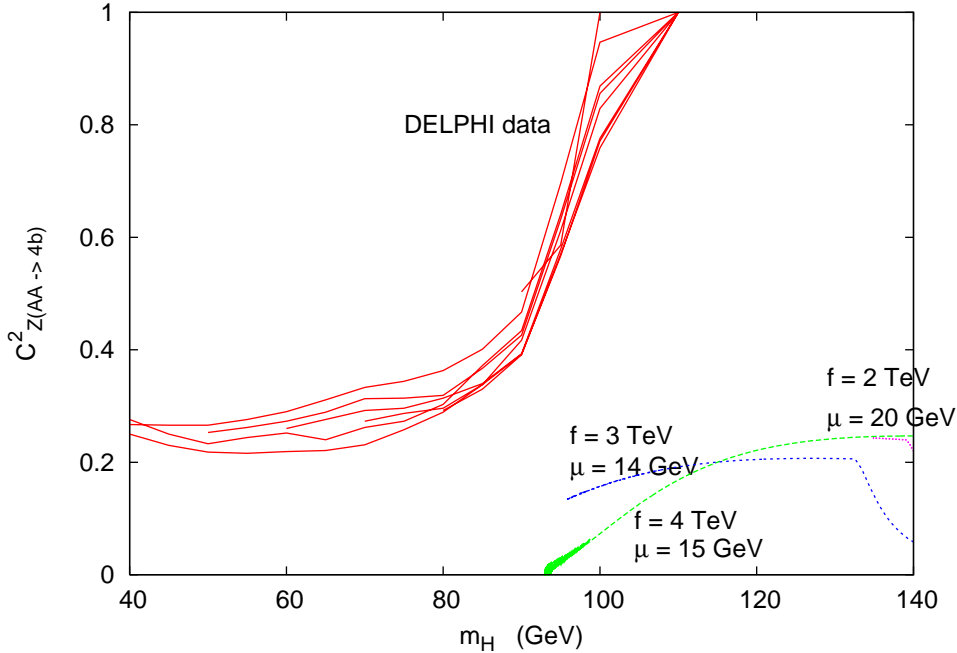}
\end{center}
    \caption {
Upper bound on $C^2_{Z(AA\to4b)}$ by the DELPHI collaboration,
and the values of $C^2_{Z(AA\to4b)}$ in our model for $f=2,3,4$ TeV.
    }
    \label{fig:delphi}
\end{figure}

We show the values of $C^2_{Z(AA\to4b)}$ predicted in
our model for $f=2,3,4$ TeV in Fig.~\ref{fig:delphi}.
Here we fix $\mu=20,\;14,\;15\gev$ for $f=2,3,4$ TeV, respectively.
We also show the upper bounds on $C^2_{Z(AA\to4b)}$
for various combinations of $m_H$ and $m_A$ obtained
by the DELPHI collaboration.\footnote{The mass ranges of the DELPHI data are
$12 \,{\rm GeV} < m_A < 55 \,{\rm GeV}$ and
$2\, m_A < m_H < 110\,{\rm GeV}$.
} For all  three cases
the $C^2_{Z(AA\to4b)}$ values in this model are much smaller than the
experimental upper bound.
The DELPHI searches do not constrain the model at all.
For $f=2\tev$ case, it is because the Higgs boson mass is already above
the lower bound of 114.4 GeV.
For $f=3,4\tev$, smaller $m_H$ can evade the DELPHI search
since $g_{ZZH}$ decreases substantially for large $\tbt$
and $H\to b\bar{b}$ is still dominant for
$m_H\lsim 100\gev$ as discussed before.
The kinks in the curves are due to the onset of the $Z \eta$ mode when
$m_H > m_Z + m_\eta$.

\section{Conclusions}
\label{sec:Conclusions}

Little Higgs models provide a very interesting perspective on
answering the little hierarchy problem.
As attributing the lightness
of Higgs boson to its being a pseudo Nambu-Goldstone boson, the
collective symmetry breaking mechanism removes the quadratically divergent
radiative-corrections to the Higgs mass at one-loop level.
As a perfect type of ``simple group" models, the SU(3)
simplest little Higgs model has drawn a lot of interests due to its
lowest fine-tuning associated to electroweak symmetry
breaking\,\cite{fine-tuning}.  In the original framework, this simplest
model cannot avoid the presence of massless pseudoscalar particle
$\eta$.
Cosmological lower bound on the axion mass requires
to extend the model.
One of the simplest choices is to add the so-called $\mu$ term
in the scalar potential by hand.
Then $\eta$ acquires a mass of order $\mu$, and
the $H$-$\eta$-$\eta$ coupling is also generated of the order of $v\mu^2/f^2$.
In order to accommodate the EWSB,
this $\mu$ has a natural scale of a few ten GeVs,
which leads to relatively light $\eta$.
It is possible to
allow a substantial branching ratio for the $H \to \eta\eta$ decay.
In addition,
the $H$-$Z$-$\eta$ coupling, which is present in the original model without the $\mu$ term,
leads to $H \to Z\eta$ decay.

We found that the $H\to \eta\eta$ decay can be dominant for $m_H$
below the $WW$ threshold for $\mu \simeq 15-20$ GeV, while
$H \to Z \eta$ dominant if $140 \gev \lsim m_H \lsim 2 m_W$.
For $m_H$ even above $2 m_W$,
the $H \to Z \eta$ decay can be as important as $H \to W^+ W^-$.
We have investigated the LEP bound on $[g_{ZZH}/g^{\rm SM}_{ZZH}]^2
B(H \to b \bar{b})$
in the search for the SM Higgs boson.
In the $f=2\tev$ case,
the model restricts $m_H$ above the LEP bound.
For the $f=3\,(4)\tev$ cases,
a lowering in the Higgs boson mass bound occurs:
$m_H > 109\; (111)\gev$, respectively.  This is the main result of our work.

A few comments are in order here.

\begin{itemize}

\item
This new and dominant decay channel can lead to important implications
on the LEP search for the neutral Higgs boson.
The DELPHI collaboration examined, in extended models, the process of
$e^+ e^- \to H Z \to (AA) Z \to (b \bar{b} b \bar{b})Z$,
and presented the upper bound on
$[g_{ZZH}/g^{\rm SM}_{ZZH}]^2 B(H\to\eta\eta) B(\eta\to b \bar{b})^2 $.
Our models with $f=2,3,4\tev$ are not constrained by this bound.
%In the $f=2\tev$ case, the minimum of the Higgs boson mass
%is rather heavy, above the LEP bound 114 GeV;
%in the $f=3,4\tev$ case, large $\tbt$ suppresses the coupling $g_{ZZH}^2$.

%\item
%Since the $\eta$ boson is rather light, one may suspect that it should
%have been found in decays of Upsilon, $J/\psi$, and some other mesons.
%Considering the lower bound on CP-odd scalar
%from $b$-physics around $\mathcal{O}(100)$ MeV,
%this $\eta$ with generical mass scale of $10\gev$
%is safe.

\item
Further probes of the scenario are possible at LEP, at the Tevatron, and
at the LHC. The LEP collaborations can investigate the scenario
by searching for
\[
  e^+ e^- \to ZH \to Z\,(\eta \eta) \to Z ( 4b,\, 2b\, 2\tau, 4\tau) \;,
\]
where $Z \to \ell^+ \ell^-,\, \nu\bar \nu,\,q\bar q$.  This mode may suffer
from the fact that the coupling $g_{ZZH}$ is reduced relative to the SM
one because of the little Higgs corrections.
At the Tevatron, similar channels such as
\[
 p \bar p \to W H, ZH \to W/Z + (4b,\, 2b \,2\tau,\, 4 \tau)
\]
can be searched for.
At the LHC, the two-photon decay mode of the intermediate Higgs boson will
suffer because of the dominance of $H \to \eta \eta$ mode in that mass
range.  Thus, the branching ratio into $\gamma\gamma$ reduces.  On
the other hand, $gg \to H \to \eta \eta \to 4b,\, 2b \,2\tau ,\, 4 \tau$ open,
which may be interesting modes to search for the Higgs boson.  However,
a detailed study is needed to establish the feasibility.

\item The $Z \eta$ decay mode of the Higgs boson is very unique in this simplest
little Higgs model.
%As long as $m_H > m_Z + m_\eta$, this mode can be quite important.
In fact it
dominates for $140 \;{\rm GeV} < m_H < 2 \, m_W$.  Even for $2 \, m_W < m_H$ the
$Z \eta$ mode is as important as $WW$ mode.
It is very different from a SM-like Higgs boson, which usually has the
$ZZ$ mode in the second place.
Since the $ZZ \to \ell^+ \ell^- \ell^+ \ell^-$ is the golden mode for
Higgs discovery, the emergence of the $Z \eta$ mode will affect the
Higgs detection significantly.  Careful studies of $Z \eta$ mode is
therefore important for Higgs searches.

\item
Another possibility to probe the $\eta$
is the direct production of the $\eta$ boson in $gg$
fusion \cite{SingleT} or the associated production with a heavy quark
pair.  Although the production is suppressed by $1/f$ in the coupling
of the $\eta$ to the SM fermion pair, this remains as an interesting possibility
because the coupling to the heavy top quark is not suppressed.

\end{itemize}

We end here with an emphasis that $4b, 2b\, 2\tau, 4\tau$ modes should be
seriously searched for in the pursuit of the Higgs boson, which we
have clearly demonstrated that it is possible in the simplest little
Higgs models for $H\to \eta \eta$ and $H \to Z \eta$ to be dominant.

\acknowledgments

We thank the Physics division of the KIAS for hospitality during the
initial stage of the work.
K.C. also thanks K.S. Cheng and the Centre of Theoretical and Computational
Physics at the University of Hong Kong for hospitality.
And we would like to express our special gratitude to Alex G. Dias,
for correcting our mistakes.
We also appreciate the valuable comment from Juergen Reuter.
The work of JS is supported by KRF under grant No. R04-2004-000-10164-0.
The work of KC is supported by
the National Science Council of Taiwan under grant no.
95-2112-M-007-001- and by the National Center for Theoretical Sciences.

\end{document}